\documentclass[12pt,twocolumn]{article}
\begin{document}
\title
{Dirac equation in a gauge-field background in the Moyal plane} 
\author{
{\bf {\normalsize Aslam Halder}$^{a}
$\thanks{aslamhalder.phy@gmail.com}},
{\bf {\normalsize Sunandan Gangopadhyay}$^{b}$\thanks{sunandan.gangopadhyay@gmail.com}},\\[0.2cm]
{\bf {\normalsize Anirban Saha}$^{a}$\thanks{anirban@wbsu.ac.in}}\\[0.2cm]
$^{a}${\normalsize Department of Physics, West Bengal State University, Barasat, Kolkata 700126, India}\\
$^{b}$ {\normalsize  Department of Astrophysics and High Energy Physics,}\\
{\normalsize S. N. Bose National Centre for Basic Science,} \\
{\normalsize Block-JD, Sector-III, Salt Lake, Kolkata-700106, India}\\[0.2cm]
}

\date{}

\maketitle
\begin{abstract}
Starting with the Dirac equation for an electron in a constant electromagnetic background on a noncommutative (NC) plane, we obtain a gauge invariant description of the system. Surprisingly, the dynamics of the system is dictated by the standard form of Lorentz force law, once the effective magnetic and electric fields $\left( B^{NC},  \, E^{NC} \right)$ correct up to leading order in the NC parameter are identified. The Hall effect is studied using the NC corrected fields in the non-relativistic (NR) limit. This shows that noncommutativity affects the cyclotron frequency, but leaves the Hall conductivity unaffected at least to first order in the NC parameter. Owing to the NC corrected magnetic field, the hyperfine splitting of Hydrogen atom spectrum also shows a first order correction which helps establish an upper bound on the spatial NC parameter.
\end{abstract}

\maketitle

\vskip 1cm
\section{Introduction}

In the last two decads systems living in a noncommutative (NC) space has drawn a lot of interest \cite{sw}-\cite{jellal billati}. From string theory with $D$-branes in a Neveu-Schwarz field background \cite{acny}-\cite{bcsgscholtz} one can arrive at a low energy effective field theory in the point particle limit,  yielding a noncommutative (NC) quantum field theory (QFT) \cite{sw}, \cite{scho}-\cite{szabo}, \cite{chams}-\cite{sghazra}. Various quantum gravity theories also lead to NC geometry \cite{suss}-\cite{mof}. In NCQFT the NC coordinate algebra $\left[\hat{X}^{\mu},\hat{X}^{\nu}\right]=i\theta^{\mu\nu}$, with the constant anti-symmetric tensor $\theta^{\mu\nu}$, creates uncertainty among the space-time coordinates. Hence the notion of a space-time point gets replaced by a cell. Interestingly, in complete analogy with the stringy scenario, noncommuting coordinates also arise in a simple quantum mechanical setting like the Landau problem where a charged particle moves in an electromagnetic background, e.g., in quantum Hall effect with partially filled lowest Landau Level.  
Therefore, it is not surprising that the study of a charged particle living in the NC plane with a constant electromagnetic background has had considerable emphasis in the literature \cite{duval}-\cite{gov}. Naturally, these studies are performed at the low energy limit of NCQFT,  i.e., using NC quantum mechanics (QM).  

 In the literature a plethora of such studies are available in the non-relativistic regime \cite{duval}-\cite{ahsg2}. Some authors \cite{dmgitman}-\cite{kai} also aproach the system by starting in the relativistic domain and eventually taking the NR limit. In these studies the issue of NC gauge invariance often goes unaddressed \cite{dmgitman,  bertolami} despite the presence of a gauge-field background. This leads to a menifestly non-gauge-invariant theory. For example, in \cite{bertolami}, the effect of spatial noncommutativity could not be incorporated because that would render the Dirac Hamiltonian non-gauge-invariant, resulting in a gauge-dependent expression for the particle's velocity. Our present work fixes this problem by taking a menifestly gauge-invariant approach.

We address the issue of gauge-dependence right at the outset, by starting with a field theoretic action where ordinary product is replaced by Moyal star $\left( \star-\right)  $product \cite{szabo} and ordinary fields are replaced by their NC counterparts to arrive at the NC field theory. This NC theory can be expressed in terms of commutative field variables by perturbatively expanding \cite{mezin} the $\star-$product and subsequently using the Seiberg-Witten (SW) map \cite{sw} that relates the family of NC gauge fields connected by the NC gauge transformation to ordinary gauge fields connected by standard gauge transformation. We refere to this as the {\it commutative equivalent} description \cite{carroll,  bcsgas} and since it yields a theory not only manifestly gauge-invariant but also in terms of commutative field variables with NC correction terms, physical interpretation of quantities like particle velocity and force are natural and straightforward.

The organization of the paper is as follows: In section 2 we demonstrate how simply replacing the ordinary product in the quantum mechanical Hamiltonian of the system by $\star-$product fails to generate a proper NC quantum mechanical Hamiltonian that can preserve the gauge invariance. In section 3 we resolve this issue by upgrading to a NC field theoretic description.
We elaborate how one can arrive at the Hamiltonian in the commutative equivalent description that takes into account NC effects while keeping the gauge invariance intact. This Hamiltonian dictates the time evolution yielding the Lorentz force law in its standard form but with NC corrected magnetic and electric fields. The non-relativistic limit of the NC Dirac Hamiltonian is used to study the Hall effect in section 4 and the hyperfine splitting of Hydrogen atom (H-atom) spectrum in section 5. We summarize in section 6.

\section{Issue with NC generalization of Dirac equation:}
In this section let us demonstrate the inherent problem with the strategy of generalising a standard quantum mechanical system to its NC counterpart at the equation of motion level, if the system has a gauge field. Consider an electron of mass $m$ moving on a two dimensional NC plane in the presence of a background EM field. We want to study its relativistic dynamics quantum mechanically, so we attempt to extend the standard Dirac equation for such an electron as
\begin{eqnarray}
\label{evx}
i\hbar\frac{\partial\Psi(X)}{\partial t}=\hat{H}(X, P)\star\Psi(X)
\end{eqnarray}
with the NC Dirac Hamiltonian of the electron $\hat{H}$ given by
\begin{eqnarray}
\hat{H}=c\vec{\alpha}\cdot\left(\vec{\hat{P}}-\frac{e}{c}\vec{\hat{\mathcal{A}}}\left(X\right)\right)+\beta mc^2+e {\hat \phi }\left(X\right)
\label{NC_Ham}
\end{eqnarray}
where ${\hat{\mathcal{A}}}_{\mu} = \left({\hat \phi }\left(X\right),  \,  - \vec{\hat{\mathcal{A}}}\left(X\right)\right)$ is NC gauge field, $\Psi\left(X\right)$ is the NC spinor of the electron and $\alpha$, $\beta$ are the well known Dirac matrices. The key idea here is to replace the ordinary product among the functions of the commutative variables $f(x)$ and $g(x)$ by the Moyal $-\star$ product defined as
\begin{eqnarray}
f(X)\star g(X)=f(x)exp\left\{\frac{i}{2}\theta^{\mu\nu}\overleftarrow{\partial_{\mu}}\overrightarrow{\partial_{\nu}}\right\}g(x)
\label{ev03}
\end{eqnarray}
among the corresponding functions of the NC variables $f(X)$ and $g(X)$.
Applying the definition (\ref{ev03}) to first order in spatial NC parameters $\theta^{ij}$ in eq.(\ref{evx}) we get the modified Dirac equation as
\begin{eqnarray}
\label{ej467}
i\hbar\frac{\partial\Psi(x)}{\partial t}&=&\hat{H}(x, p)\Psi(x)+\frac{i}{2}\theta^{jk}\partial_{j}\hat{H}(x, p)\partial_{k}\Psi(x)\nonumber\\
&=&\hat{H}(x, p)\Psi(x)+\frac{i}{2}\theta^{jk}\partial_{j}\left\{c\alpha_{j}\left(\hat{p}_{j}-\frac{e}{c}\hat{\mathcal{A}}_{j}(x)\right)\right.\nonumber\\
&&\left.+\beta mc^2+e\phi\right\}\partial_{k}\Psi(x)~.
\end{eqnarray}
From (\ref{ej467}) we can read off the NC Dirac Hamiltonian and with a slight rearrangement of notations, e.g., $\theta^{i}=\frac{1}{2}\epsilon_{ijk}\theta^{jk}$, it can be writen as
\begin{eqnarray}
\label{es04llb}
\hat{H}&=&c\vec{\alpha}\cdot\left(\vec{\hat{p}}-\frac{e}{c}\vec{\hat{\mathcal{A}}}(x)\right)+\beta mc^2+e\phi\nonumber\\
&&+\frac{e}{2\hbar}\left\{\bigtriangledown\left(c\vec{\alpha}\cdot\vec{\hat{\mathcal{A}}}-\phi\right)\times\vec{\hat{p}}\right\}\cdot\vec{\theta}~.
\end{eqnarray}
where it is evident that the NC correction terms spoil the gauge-invariance. Therefore this Hamiltonian is not suitable for studying the dynamics of a relativistic electron. In \cite{bertolami} this issue was bypassed by discarding the spatial noncommutativity totally which rendered what was a very important analysis otherwise, incomplete.

As will be elaborated in the next section, we tackle this issue of gauge-non-invariance by implimenting the NC generalization at the action level instead and arriving at a {\it commutative equivalent description} \cite{carroll, bcsgas} of the NC system. Similar approach was undertaken in \cite{Adorno} and \cite{kai}  to study the effect of noncommutativity in Pauli equation and quantum phases respectively.

\section{NC Dirac field theory}
We start by generalizing the commutative Dirac field action of an electron moving in a electromagnetic background to the NC space. We follow the same key idea of redefining the field variables and their product rule as stated in the previous section. This brings us to the U$(1)_{\star}$ gauge invariant action for the NC spinor field ${\Psi}(X)$ coupled with a background U$(1)_{\star}$ gauge field ${\mathcal{A}}_{\mu}\left( X \right) = ({\mathcal{A}}_{0}, {\mathcal{A}}_{i}),~i=1,2,3;$ 
\begin{eqnarray}
S=\int d^4x ~{\Psi}^{\dagger}\star \left(\gamma^{\mu}\Pi_{\mu} - mc \right) \star{\Psi}  
\label{eur}
\end{eqnarray}
where
\begin{eqnarray}
\Pi_{\mu}=P_{\mu} -\frac{e}{c}\mathcal{A}_{\mu}\left( X \right)
\label{NC_Mech_momentum}
\end{eqnarray}
is the NC version of the standard mechanical momentum 
\begin{eqnarray}
\pi_{\mu} = p_{\mu} -\frac{e}{c} A_{\mu} \left( x \right)
\label{Mech_momentum}.
\end{eqnarray}
To get to the commutative equivalent picture we expand the $\star-$ product in the action (\ref{eur}) to first order and express the NC gauge field ${\mathcal{A}}_{\mu}\left( X \right)$ and the NC spinor field ${\Psi}\left( X \right)$ in terms of their ordinary commutative counterparts $A_{\mu}\left( x \right)$ and  $ \psi(x) $ using the relevant SW maps \cite{sw},  which to first order in the NC parameters, read
\begin{eqnarray}
{\mathcal{A}}_{\mu} & = & {A}_{\mu}+\frac{e}{2\hbar c}\theta^{\alpha\beta}{A}_{\alpha} \left( \partial_{\beta}{A}_{\mu}+F_{\beta\mu} \right) \label{e159} \\
{\Psi}& = &\psi+\frac{e}{2\hbar c}\theta^{\alpha\beta}{A}_{\alpha}\partial_{\beta}\psi 
\label{SW_Psi}~.
\end{eqnarray}
This yields the action \cite{Adorno}
\begin{eqnarray}
S^{\theta}&=&\int d^4x~{\psi}^{\dagger}(x)\left\{\gamma^{\mu}\left[\left(1+\frac{e}{4c\hbar}\theta^{\alpha\beta}F_{\alpha\beta}\right)\pi_{\mu}
\right. \right. \nonumber\\&& \left. \left.-\frac{e}{2c\hbar}\theta^{\alpha\beta}F_{\alpha\mu}\pi_{\beta}\right]\right.\nonumber\\&& \left.-mc\left(1+\frac{e}{4c\hbar}\theta^{\alpha\beta}F_{\alpha\beta}\right)\right\}\psi(x)~.
\label{e163vx}
\end{eqnarray}
Owing to the appearance of U$\left(1\right)$ gauge-invariant field strength tensor 
\begin{eqnarray}
F_{\alpha\beta}&=&\partial_{\alpha}A_{\beta}(x)-\partial_{\beta}A_{\alpha}(x)
\label{NC_field_strength}
\end{eqnarray}
of standard electrodynamics and commutative mechanical momentum $\pi$,  defined in (\ref{Mech_momentum}), this action is manifestly gauge invariant.  Let us reinforce that employing the SW map along with the $\star-$ product ensures that we get  an action that depicts the NC system in terms of the commutative variables while keeping the gauge-invariance intact.  

Varying the action (\ref{e163vx}) we obtain the equation of motion which upon quantization takes the same form as the Dirac equation in relativistic quantum mechanics (RQM)
\begin{eqnarray}
\label{e163x9}
i\hbar\partial_{t}\psi&=&\hat{{H}}_{{\rm NC}}\psi
\end{eqnarray}
but with the NC-corrected Hamiltonian operator
\begin{eqnarray}
\hat{H}_{{\rm NC}}&=&c\vec{\alpha}\cdot\left(\vec{\hat{p}}-\frac{e}{c}\vec{\hat{A}}\right)\left(1+\frac{eB\theta}{2\hbar c}\right)-eE\hat{x}\nonumber\\
&&+\frac{eE\theta}{2\hbar}\left(\hat{p}_{y}-\frac{e}{c}\hat{A}_{y}\right)+\beta mc^2\nonumber\\
&=&c\left\{(\alpha_{x}\hat{p}_{x}+\alpha_{y}\hat{p}_{y})+\frac{eB}{2c}(\alpha_{x}\hat{y}-\alpha_{y}\hat{x})\right\}\nonumber\\
&&\times\left(1+\frac{eB\theta}{2\hbar c}\right)-eE\hat{x}+\frac{eE\theta}{2\hbar}\left(\hat{p}_{y}-\frac{eB\hat{x}}{2c}\right) \nonumber\\
&&+\beta mc^2.
\label{e163kj}
\end{eqnarray}
In writing (\ref{e163kj}) we have chosen the direction of electric field and magnetic field\footnote{We have taken the symmetric gauge,  $A_{x}=  -\frac{By}{2},  A_{y}=  \frac{Bx}{2}$.} as $\vec{E}=E\hat{i}$ and $\vec{B}=B\hat{k}$ and also used the notation $\theta_{ij} = \epsilon_{ij} \theta, \, \left( i,j = 1,2\right)$ since the electron under consideration is confined to the NC $x-y$ plane. To avoid any issue with the causality we have discard the time-space noncommutativity by setting $\theta^{\mu0}=0$ as is the usual practice.
In contrast with the Hamiltonian (\ref{es04llb}),  $H_{{\rm NC}}$ is manifestly gauge invariant. This is only natural since it came from the gauge-invariant action (\ref{e163vx}).

The Hamiltonian (\ref{e163kj}) dictates the time evolution of the momentum operator which, with the usual operator correspondence for velocity $\vec{\hat{v}} = c\vec{\alpha}$, can be cast in the form of a Lorentz force equation  
\begin{eqnarray}
\label{e04xgy} 
\dot{\vec{\hat{\pi}}}&=&\frac{d\vec{\hat{\pi}}}{dt}\nonumber\\ 
&=&\frac{i}{\hbar}[\hat{H},  \vec{\hat{\pi}}]\nonumber\\
&=&\frac{i}{\hbar}\left[c\left\{(\alpha_{x}\hat{p}_{x}+\alpha_{y}\hat{p}_{y})+\frac{eB}{2c}(\alpha_{x}\hat{y}-\alpha_{y}\hat{x})\right\}\right.\nonumber\\
&&\left.\left(1+\frac{eB\theta}{2\hbar c}\right)-eE\hat{x}+\frac{eE\theta}{2\hbar}\left(\hat{p}_{y}-\frac{eB\hat{x}}{2c}\right)\right.\nonumber\\
&&\left.+\beta mc^2~,~\hat{i}\left(\hat{p}_{x}-\frac{e}{c}\hat{A}_{x}\right)+\hat{j}\left(\hat{p}_{y}-\frac{e}{c}\hat{A}_{y}\right)\right]\nonumber\\
&=&e\left(\frac{1}{c}\vec{\hat{v}}\times\vec{B}+ \vec{E}\right)\left(1+\frac{eB\theta}{2\hbar c}\right)
\end{eqnarray}
where the Lorentz force percived by the electron is seen to pick up a NC correction. From (\ref{e04xgy}) it is evident that either the electromagnetic fields or the charge carried by the electron can be viewed to have been modified by the spatial noncommutativity, we choose to go with the formar option\footnote{Note that going with the other option of a NC modified electric charge would lead to the same physical results.} and define the NC corrected electric and magnetic fields as
\begin{eqnarray}
\vec{E}^{NC}=\vec{E}\left(1+\frac{eB\theta}{2\hbar c}\right)\nonumber \\
\vec{B}^{NC}=\vec{B}\left(1+\frac{eB\theta}{2\hbar c}\right)
\label{e04xycs}
\end{eqnarray}
to write the NC Lorentz force as
\begin{eqnarray}
\label{e04xgy1} 
\vec{F}^{NC} = \frac{d\vec{\hat{\pi}}}{dt} = \frac{1}{c}e\vec{\hat{v}}\times \vec{B}^{NC}+e\vec{E}^{NC}~.
\end{eqnarray}
However in the non-relativistic limit of the Dirac equation\footnote{Shown in Appendix 1.} one can identify the corresponding Hamiltonian 
\begin{eqnarray}
\label{e0123x}
\hat{H}_{Pauli}^{NC}&=&\frac{1}{2m}\left(\vec{\hat{p}}-\frac{e}{c}\vec{\hat{A}}\right)^{2}\left(1+\frac{eB\theta}{\hbar c}\right)\nonumber\\&
&-\frac{e\hbar}{2mc}\vec{\sigma}\cdot\vec{B}
\left(1+\frac{eB\theta}{\hbar c}\right)
\end{eqnarray}
where the first term dictates the time evolution of the canonical degrees of freedom (DOF) leading to the same Lorentz force equations but with the different NC corrected fields as we shall see in the next section and the second term shows the spin DOF,  a relic of the relativistic effect.   
In the reminder of the paper we shall use the NR Hamiltonian (\ref{e0123x}) in context of Hall effect and hyperfine splitting of Hydrogen atom to see how spatial noncommutativity affacts an electron's perception of electromagnetic fields.  
%
\section{Cyclotron motion and Hall conductivity in the Moyal plane}
In usual quantum mechanics the electron on a plane with a uniform perpendicular magnetic background leads to cyclotron motion and we can easily compute the Hall conductivity in this scenario using the Streda formula \cite{streda} once we obtain the degeneracy of the quantum state of the electron. We shall apply the algorithm presented in \cite{yan} with our commutative equivalent NC non-relativistic Hamiltonian (\ref{e0123x}) to compute the said degeneracy when the electron moves in a NC plane.

The time evolution of the position and velocity components can be obtained using the Ehrenfest theorem  
\begin{eqnarray}
\dot{x} &=& \frac{i}{\hbar} \langle \left[\hat{H}, \hat{x}\right]\rangle\nonumber\\
 &=& \frac{1}{m}\left(p_{x} + \frac{eBy}{2}\right)\left(1+\frac{eB\theta}{\hbar c}\right)
\label{x_dot_NC} \\
\dot{y} &=& \frac{i}{\hbar} \langle \left[\hat{H}, \hat{y}\right]\rangle\nonumber\\
&=&\frac{1}{m}\left(p_{y} - \frac{eBx}{2}\right)\left(1+\frac{eB\theta}{\hbar c}\right)
\label{y_dot_NC} 
\end{eqnarray}
and 
\begin{eqnarray}
\ddot{x} &=& \frac{i}{\hbar}\langle \left[\hat{H}, \dot{\hat{x}}\right]\rangle=\left(1+\frac{eB\theta}{\hbar c}\right)\left(\frac{eB}{mc}\right) \dot{y}\nonumber\\
&=&\left(\frac{eB^{\rm{NC}}}{mc}\right) \dot{y}
\label{x_ddot_NC} \\
\ddot{y} &=& \frac{i}{\hbar}\langle \left[\hat{H}, \dot{\hat{y}}\right]\rangle= - \left(1+\frac{eB\theta}{\hbar c}\right)\left(\frac{eB}{mc}\right) \dot{x}\nonumber\\
  &=& -\left(\frac{eB^{\rm{NC}}}{mc}\right) \dot{x}
\label{y_ddot_NC} 
\end{eqnarray}
where we can readily see the magnetic Lorentz force $\frac{e}{c}\left(\vec{v}\times\vec{B}^{\rm{NC}} \right)$ in the second set of equations, however the NC correction factor in the magnetic field $\vec{B}^{\rm{NC}}=\vec{B}\left(1+\frac{eB\theta}{\hbar c}\right)$ is different from that of the relativstic case in eq. (\ref{e04xycs}).

\noindent The first integral of the magnetic force equations yields 
\begin{eqnarray}
\dot{x}&=& \tilde{\omega}_{{\rm{c}}}\left(y - y_{0}\right)  = \tilde{\omega}_{{\rm{c}}}R_{y}\nonumber \\
\dot{y}&=& \tilde{\omega}_{{\rm{c}}}\left(x - x_{0}\right) = \tilde{\omega}_{{\rm{c}}} R_{x}
\label{emsa}
\end{eqnarray}
where  $\tilde{\omega}_{{\rm{c}}}=\frac{eB^{\rm{NC}}}{mc}$ is the NC corrected cyclotron frequency,  $\vec{R} = \left(x - x_{0}\right)\hat{i} + \left(y - y_{0}\right)\hat{j}$ is the radius of the cyclotron motion around the center $\left( x_{0},  \,y_{0}\right)$.   
Since these center of motion coordinates are the constants of motion, we can compute their commutator using equations (\ref{emsa}) and (\ref{x_dot_NC}, \ref{y_dot_NC}) as
\begin{eqnarray}
\label{emsuy1}
\left[\hat{x}_{0},  \hat{y}_{0} \right]&=& \left[\hat{x} + \frac{1}{\tilde{\omega}_{c}} \dot{\hat{y}} , \,\,\,  \hat{y} - \frac{1}{\tilde{\omega}_{c}} \dot{\hat{x}} \right] = - \frac{i \hbar c}{eB}\nonumber\\
 &=& -i l_{{\rm{B}}}^2
\end{eqnarray}
where $l_{{\rm{B}}}=\sqrt{\frac{\hbar c}{eB}}$ is interpreted as the magnetic length characterizing the length scale of the system.  
This nonvanishing commutator (\ref{emsuy1}) signifies that there is a Heisenberg uncertainty associated with the coodinates of the cyclotron motion center.  The centre is,  therefore,  smeared out over a surface area $\Delta x_{0}\Delta y_{0}=2\pi l_{{\rm{B}}}^2$. Since the magnetic length scale $l_{{\rm{B}}}$ in (\ref{emsuy1}) does not pick up any NC correction this minimal area of uncertainty is unaffected by noncommutativity.  

Inverse of this minimal area gives the number of quantum mechanical states allowed per unit area or the flux density 
\begin{eqnarray}
n_{\rm{B}} = \frac{1}{2\pi l_{{\rm{B}}}^2} = \frac{eB}{hc}~.
\label{flux_density}
\end{eqnarray}
The electron density $n_{\rm{e}}$ per unit flux density $n_{\rm{B}}$ defines the filling factor $\nu = \frac{n_{\rm{e}}}{n_{\rm{B}}}$, which is, in turns,  used to express the electron density in the Streda formula
\begin{eqnarray}
\label{e9jd}
\sigma_{\rm{H}} = e\frac{\partial n_{{\rm{e}}}}{\partial B} = e \nu \frac{\partial n_{{\rm{B}}}}{\partial B} = \frac{e^2\nu}{hc}
\end{eqnarray}
to compute the Hall conductivity $\sigma_{\rm{H}}$. Thus we see the Hall conductivity remains unaffected by spatial noncommutativity. This result is in tune with the findings of \cite{bcsgas}. However this is contrast with \cite{harms}, where the Hall conductivity is modified by the spatial NC parameter $\theta$ in the leading order.  
 
The trajectory of the cyclotron motion in the NC plane can be easily obtained once we use (\ref{emsa}) to substitute for $\dot{x}$ and $\dot{y}$ in (\ref{x_ddot_NC},  \ref{y_ddot_NC})  and use $\ddot{R}_{x} = \ddot{x}$ etc:
\begin{eqnarray}
\label{eokh}
x(t)&=&x_{0}+R \sin(\tilde{\omega}_{{\rm{c}}}t+\phi)\\
y(t)&=&y_{0}+R \cos(\tilde{\omega}_{{\rm{c}}}t+\phi)\nonumber
\end{eqnarray} 
which reveals that but for the the cyclotron frequency $\tilde{\omega_{{\rm{c}}}} = \frac{eB^{\rm{NC}}}{mc}$, the cyclotron motion is also largely unaltered by the spatial noncommutativity. 

\section{Noncommutative effect in the hyperfine splitting of Hydrogen atom spectrum}
Hyperfine splitting results from the interaction of the nuclear magnetic field with the magnetic moment of orbital electron. In the case of Hydrogen atom the hyperfine splitting can be well explained by incorporating the relativistic correction into the Hamiltonian. To obtain the same one has to make the Foldy-Wouthuysen transformation \cite{greiner} of the Dirac Hamiltonian of a relativistic electron, moving in the presence of a magnetic field $B$, yielding the non-relativistic Hamiltonian where the effect of relativity results in a spin-dependent perturbation term  
\begin{eqnarray}
\label{e104gx} 
\hat{H}_{\rm{hfs}}=\frac{e\hbar}{2mc}\vec{\sigma}\cdot\vec{B}
\end{eqnarray}
which is responsiable for the hyperfine splitting. Here $\sigma$ is the Pauli spin matrix of electron. 
The corresponding energy correction is 
\begin{eqnarray}
\label{e14gx} 
E_{\rm{hfs}}=<\psi_{nl}|\hat{H}_{{\rm{hfs}}}|\psi_{nl}>
\end{eqnarray}
where $\psi_{nl}$ is the unperturbed wavefunction of the electron in n-th state.
This correction breaks the $j$-degeneracy by splitting the energy of singlet and triplet states (lift the triplet configuration and depressed the singlet). 
For the ground state of Hydrogen atom (i.e $1S_{1/2}$ state) we have the following hyperfine energy corrections in commutative space \cite{grifith}
\begin{eqnarray}
\label{e14gmm13} 
<\psi_{10}|\hat{H}_{\rm{hfs}}|\psi_{10}>&=&\frac{4g_{p}\hbar^4}{3m_{p}m^2c^2a^4}\frac{1}{4}\nonumber\\
&&\quad(\rm{for~triplet~state})\nonumber\\
&=&-\frac{4g_{p}\hbar^4}{3m_{p}m^2c^2a^4}\frac{3}{4}\\
&&\quad(\rm{for~singlet~state})\nonumber
\end{eqnarray}
where $m_{p}$ and $g_{p}$ are the mass and the gyromagnetic ratio of proton and $a$ is the first Bohr radius. 
Therefore the energy gap between these states is given as 
\begin{eqnarray}
\label{e14gmm14} 
\triangle E_{hfs}&=&\frac{4g_{p}\hbar^4}{3m_{p}m^2c^2a^4}\frac{1}{4}-\left(-\frac{4g_{p}\hbar^4}{3m_{p}m^2c^2a^4}\frac{3}{4}\right)\nonumber\\
&=&\frac{4g_{p}\hbar^4}{3m_{p}m^2c^2a^4}\nonumber\\
&=&5.88\times 10^{-6} eV.
\end{eqnarray}
This corresponds to a transition that emits a photon with a frequency of approximately 1420 MHz.

We now want to study the hyperfine splitting of the Hydrogen atom spectrum in NC space. 
From the Hamiltonian (\ref{e0123x}) we can easily write the hyperfine perturbation term for NC space as
\begin{eqnarray}
\label{e14gx}
\hat{H}_{hfs}^{NC}&=&\frac{e\hbar}{2mc}\vec{\sigma}\cdot\vec{B}^{NC}\nonumber\\ 
&=&\frac{e\hbar}{2mc}\vec{\sigma}\cdot\vec{B}\left(1+\frac{eB\theta}{\hbar c}\right)~.
\end{eqnarray}
Therefore the energy correction due to hyperfine perturbation in NC space is given by
\begin{eqnarray}
\label{e14g14} 
E_{hfs}^{NC}&=&<\psi_{nl}|\hat{H}_{hfs}^{NC}|\psi_{nl}>\nonumber\\
&=&<\psi_{nl}|\frac{e\hbar}{2mc}\vec{\sigma}\cdot\vec{B}\left(1+\frac{eB\theta}{\hbar c}\right)|\psi_{nl}>\nonumber\\
&=&\left(1+\frac{eB\theta}{\hbar c}\right)<\psi_{nl}|\frac{e\hbar}{2mc}\vec{\sigma}\cdot\vec{B}|\psi_{nl}>\nonumber\\.
\end{eqnarray}
For the ground state of Hydrogen atom in NC space we now easily express the hyperfine energy splitting between the singlet and triplet states using eq. (\ref{e14gmm13}), eq. (\ref{e14gmm14}) and eq. (\ref{e14g14}) as
\begin{eqnarray}
\label{e43gm} 
\triangle E_{hfs}^{NC}&=&\triangle E_{hfs}\left(1+\frac{eB\theta}{\hbar c}\right)\nonumber\\
&=&\frac{4g_{p}\hbar^4}{3m_{p}m^2c^2a^4}\left(1+\frac{eB\theta}{\hbar c}\right).
\end{eqnarray} 
The above expression shows a very significant result of our work. It states that hyperfine energy correction due to spatial noncommutativity depends on the background magnetic field. The correction will grow with increasing background magnetic field.

Now we want to estimate a bound for the spatial noncommutativity parameter $\theta$. To do so we use the eq. (\ref{e43gm}). From experimental observation $\frac{\triangle E_{hfs}}{h}$ for the $1S_{1/2}$ state of Hydrogen atom is known with an accuracy of $0.0024$ Hz. Therefore from the eq. (\ref{e43gm}), we can write $\triangle E_{hfs}\times\left(\frac{eB\theta}{\hbar c}\right)\frac{1}{h}\leq 0.0024$. Putting the numerical values of the various constants we get an upper bound of spatial noncommutative parameter as $\theta\leq 7.9\times 10 ^{-29}$ m$^{2}$ for a nuclear magnetic field of $14$ Tesla. In energy scale this bound is equivalent to $\theta\leq$(130 MeV)$^{-2}$. In \cite{carroll} a bound is obtained as $\theta\leq$(10 TeV)$^{-2}$ whereas in \cite{stern} the bound is $\theta\leq$(4 GeV)$^{-2}$. Compared to these results, the bound obtained in the present work is much weaker.

\section{Conclusions}
In this paper we have studied the effect of spatial noncommutativity on the dynamics of a relativistic electron moves on a two dimensional NC plane in the presence of a background EM field. 
The approach adopted here is to transform the NC system to their equivalent commutative description by employing the Seiberg-Witten map along with the Moyal star product. The gauge invariant commutative equivalent Dirac Hamiltonian of the NC system in terms of commutative variables and the NC parameter $\theta$ is then obtained by this approach. The Lorentz force of the electron in NC space is then computed and it is observed that the result gets corrected by spatial noncommutativity up to first in $\theta$. 
We then move on to study the quantum Hall effect in NC space. We find that the Hall conductivity is free from spatial noncommutativity correction up to first order in $\theta$. While studying the NC quantum Hall effect, we find that the classical trajectory of an electron moving in a magnetic field on a NC plane is modified due to noncommutativity. 
Finally, we have studied the hyperfine splitting of Hydrogen atom spectrum in NC space. Here we compute the hyperfine energy splitting of Hydrogen atom spectrum and observe that the result gets NC corrected. Using the expression for the NC hyperfine energy splitting between the singlet and the triplet configuration for the ground state of Hydrogen atom and the experimental data, we estimate an upper bound on the NC parameter $\theta$. 
 
\section*{Appendix :}
 To get the non-relativistic limit of the NC Dirac Hamiltonian we consider the Foldy-Wouthuysen transformation. We have the Dirac Hamiltonian (fully relativistic) in NC space in the presence of a background magnetic field\footnote{Since $\theta^{\mu 0}=0$ in the present work, so we don't consider the electric field here.}
\begin{eqnarray}
\label{ex987} 
\hat{H}&=&c\vec{\alpha}\cdot\left(\vec{\hat{p}}-\frac{e}{c}\vec{\hat{A}}\right)\left(1+\frac{eB\theta}{2\hbar c}\right)+\beta mc^2\nonumber\\
&=&\hat{O}+\hat{\mathcal{E}}
\end{eqnarray}
where $\hat{O}=c\vec{\alpha}\cdot\left(\vec{\hat{p}}-\frac{e}{c}\vec{\hat{A}}\right)\left(1+\frac{eB\theta}{2\hbar c}\right)$ is the odd part and $\hat{\mathcal{E}}=\beta mc^2$ is the even part of the above Hamiltonian. An operator which connects the upper and lower components of the Dirac spinor $\Psi$ is called odd operator, like $\alpha$, $\gamma_{i}$, $\gamma_{5}$; where as which does not have this property is classified as even operator, like $\beta$, $\sigma$. 
The Foldy-Wouthuysen transformation is basically a unitary transformation which eliminates all odd operators from the Dirac Hamiltonian.
We introduce the canonical transformation $\Psi^{(1)}=e^{i\hat{S}}\Psi$, $\hat{H}^{(1)}=e^{i\hat{S}}\hat{H}e^{-i\hat{S}}$ with $\hat{S}=\frac{i}{2mc^2}\beta\hat{O}$, such that
\begin{eqnarray}
\label{exy}
i\hbar\partial_{t}\Psi^{(1)}=\hat{H}^{(1)}\Psi^{(1)}~.
\end{eqnarray}
Therefore, 
\begin{eqnarray}
\label{exyz}
\hat{H}^{(1)}&=&\left(1+\frac{i\hat{S}}{1!}+\frac{(i\hat{S})^2}{2!}+\cdots\right)\hat{H}\nonumber\\
&&\times\left(1+\frac{(-i\hat{S})}{1!}+\frac{(-i\hat{S})^2}{2!}+\cdots\right)\nonumber\\
&=&\hat{H}-\hat{H}i\hat{S}+\frac{i^2}{2!}\hat{H}\hat{S}^2+i\hat{S}\hat{H}-i^2\hat{S}\hat{H}\hat{S}\nonumber\\
&&+i\hat{S}\hat{H}\frac{i^2}{2!}\hat{S}^2+\frac{i^2}{2!}\hat{S}^2\hat{H}+\frac{i^2}{2!}\hat{S}^2\hat{H}(-i\hat{S})\nonumber\\
&&+\cdots\nonumber\\
&=&\hat{H}+i[\hat{S},\hat{H}]+\frac{i^2}{2!}\left[\hat{S},[\hat{S},\hat{H}]\right]\nonumber\\
&&+\frac{i^3}{3!}\left[\hat{S},[\hat{S},[\hat{S},\hat{H}]]\right]+\cdots~.
\end{eqnarray}
We want to expand $\hat{H}^{(1)}$ up to order $O(mc^2)^{-2}$. We simplify the above expression for $\hat{H}^{(1)}$ term by term. 
\begin{eqnarray}
\label{exyz2} 
[\hat{S},\hat{H}]
&=&\frac{i}{2mc^2}\beta\hat{O}(\hat{O}+\beta mc^2)\nonumber\\
&&-(\hat{O}+\beta mc^2)\frac{i}{2mc^2}\beta\hat{O}\nonumber\\
&=&i\hat{O}-\frac{i}{mc^2}\beta\hat{O}^2\nonumber.
\end{eqnarray} 
Therefore \[i[\hat{S},\hat{H}]=-\hat{O}+\frac{1}{mc^2}\beta\hat{O}^2\].  Similarly 
\begin{eqnarray*}
\frac{i}{2!}\left[\hat{S},[\hat{S},\hat{H}]\right]&=&-\frac{1}{2mc^2}\beta\hat{O}^2-\frac{1}{2m^2c^4}\hat{O}^3 \nonumber\\
\frac{i^3}{3!}\left[\hat{S},[\hat{S},[\hat{S},\hat{H}]]\right]&=&\frac{1}{6m^2c^4}\hat{O}^3-\frac{1}{6m^3c^6}\beta\hat{O}^4\nonumber\\
\frac{i^4}{4!}\left[\hat{S}[\hat{S},[\hat{S},[\hat{S},\hat{H}]]]\right] &\approx & \frac{1}{24m^3c^6}\beta\hat{O}^4\nonumber\\
\end{eqnarray*}
Adding all the terms we have
\begin{eqnarray}
\label{exyz}
\hat{H}^{(1)}&=&\hat{O}+\beta mc^2-\hat{O}+\frac{1}{mc^2}\beta\hat{O}^2-\frac{1}{2mc^2}\beta\hat{O}^2\nonumber\\
&&-\frac{1}{2m^2c^4}\hat{O}^3+\frac{1}{6m^2c^4}\hat{O}^3-\frac{1}{6m^3c^6}\beta\hat{O}^4\nonumber\\
&&+\frac{1}{24m^3c^6}\beta\hat{O}^4\nonumber\\
&=&\beta mc^2+\frac{1}{2mc^2}\beta\hat{O}^2-\frac{1}{8m^3c^6}\beta\hat{O}^4-\frac{1}{3m^2c^4}\hat{O}^3~.\nonumber
\end{eqnarray} 
Omitting all terms with odd powers of $\hat{O}$ we get
\begin{eqnarray}
\label{exyz5}
\hat{H}^{(1)}=\beta\left(mc^2+\frac{1}{2mc^2}\hat{O}^2-\frac{1}{8m^3c^6}\hat{O}^4\right)~.\nonumber
\end{eqnarray}
Now using the following identity
\begin{eqnarray}
\label{exyz51} 
(\hat{\alpha}\cdot\vec{A})(\hat{\alpha}\cdot\vec{B})=\vec{A}\cdot\vec{B}+i\vec{\sigma}\cdot(\vec{A}\times\vec{B})\nonumber
\end{eqnarray}
we simplify the $\hat{O}^2$ term as 
\begin{eqnarray}
\label{exyz501} 
\hat{O}^2&=&\left\{\vec{\alpha}\left(\vec{\hat{p}}-\frac{e}{c}\vec{\hat{A}}\right)\left(1+\frac{eB\theta}{2\hbar c}\right)\right\}^2\nonumber\\
&=&\left(\vec{\hat{p}}-\frac{e}{c}\vec{\hat{A}}\right)^2\left(1+\frac{eB\theta}{2\hbar c}\right)^2+i\vec{\sigma}\cdot\left(\vec{\hat{p}}-\frac{e}{c}\vec{\hat{A}}\right)\nonumber\\
&&\times\left(\vec{\hat{p}}-\frac{e}{c}\vec{\hat{A}}\right)\left(1+\frac{eB\theta}{2\hbar c}\right)^2\nonumber\\
&=&\left(\vec{\hat{p}}-\frac{e}{c}\vec{\hat{A}}\right)^2\left(1+\frac{eB\theta}{\hbar c}\right)+i\vec{\sigma}\cdot i\hbar\vec{\bigtriangledown}\nonumber\\
&&\times\left(\frac{e}{c}\vec{A}\right)\left(1+\frac{eB\theta}{\hbar c}\right)\nonumber\\
&=&\left\{\left(\vec{\hat{p}}-\frac{e}{c}\vec{\hat{A}}\right)^2-\frac{e\hbar}{c}\vec{\sigma}\cdot\vec{B}\right\}\left(1+\frac{eB\theta}{\hbar c}\right)~.\nonumber
\end{eqnarray}
Neglecting the term of the order $O(mc^2)^{-3}$ and using the above expression for $\hat{O}^2$, we get  
\begin{eqnarray}
\label{exyz5re}
\hat{H}^{(1)}&=&\beta\left\{mc^2+\frac{1}{2m}\left(\vec{\hat{p}}-\frac{e}{c}\vec{\hat{A}}\right)^2\left(1+\frac{eB\theta}{\hbar c}\right)\right.\nonumber\\
&&\left.-\frac{e\hbar}{2mc}\vec{\sigma}\cdot\vec{B}\left(1+\frac{eB\theta}{\hbar c}\right)\right\}~.
\end{eqnarray}
Now we decompose the four-component spinor $\Psi^{(1)}$ into two two-component spinor $\psi$ and $\chi$ by the following representation
\begin{eqnarray}
\label{exwyz56re}
\Psi^{(1)}=\left(
\begin{array}{cc}
\psi  \\
\chi \\
\end{array}
\right)~.
\end{eqnarray}
Therefore using (\ref{exyz5re}) and (\ref{exwyz56re}), the Dirac equation (\ref{exy}) then become
\begin{eqnarray}
\label{exwyz5re}
i\hbar\partial_{t}\left(
\begin{array}{cc}
\psi  \\
\chi \\
\end{array}
\right)&=&\left(
\begin{array}{cc}
\hat{H}_{Pauli}^{NC}+mc^2&0  \\
0&-\hat{H}_{Pauli}^{NC}-mc^2 \\
\end{array}
\right)\nonumber\\
&&\times\left(
\begin{array}{cc}
\psi  \\
\chi \\
\end{array}
\right)\nonumber
\end{eqnarray}
where 
\begin{eqnarray}
\label{Pauli_H_NC}
\hat{H}_{Pauli}^{NC}&=&\frac{1}{2m}\left(\vec{\hat{p}}-\frac{e}{c}\vec{\hat{A}}\right)^2\left(1+\frac{eB\theta}{\hbar c}\right)\nonumber\\
&&-\frac{e\hbar}{2mc}\vec{\sigma}\cdot\vec{B}\left(1+\frac{eB\theta}{\hbar c}\right)
\end{eqnarray}
 is the non-relativistic Hamiltonian, also called the Pauli Hamiltonian with NC correction.




\end{document}